# Conversational Swarm Intelligence (CSI) Enhances Groupwise Deliberation

Louis Rosenberg[1], Gregg Willcox[1], Hans Schumann[1] and Ganesh Mani[2]

[1] Unanimous AI, Arlington Virginia, USA
[2] Carnegie Mellon University, Pittsburgh Pennsylvania, USA

**Abstract.** Real-time conversational deliberation is a critical groupwise method for reaching decisions, solving problems, evaluating priorities, generating ideas, and producing insights. Unfortunately, real-time conversations are difficult to scale, losing effectiveness as groups grow above 5 to 7 members. Conversational Swarm Intelligence (CSI) is a new technology modeled on the dynamics of biological swarms. It aims to enable networked groups of any size to hold productive real-time deliberations that converge on unified solutions. CSI leverages the power of Large Language Models (LLMs) in a unique and powerful way, allowing real-time dialog among small local groups while simultaneously enabling efficient content propagation across much larger populations. In this way, CSI combines the benefits of small-scale deliberative reasoning and large-scale collective intelligence. In this study, we compare deliberative groups of 48 people using standard online chat to the same sized groups using a prototype chat-based CSI system called Thinkscape™. Results show that participants using CSI contributed 51% more content (p<0.001) than those using standard chat, and the deliberations using CSI showed 37% less difference in contribution quantity between the most active vs least active members, indicating more balanced dialog. And finally, a large majority of participants preferred deliberating using the CSI system over standard chat (p<0.05) and reported feeling more impactful when doing so (p<0.01). These results suggest that Conversational Swarm Intelligence is a promising technology for enabling large-scale deliberation.

## 1 Introduction

Conversational deliberation is critical for enabling human groups to reach decisions, solve problems, generate ideas, produce insights, make plans, and prioritize objectives. Unfortunately, real-time deliberation is difficult to scale, losing effectiveness as groups grow beyond 5 to 7 members [12, 16 - 24]. As a consequence, those wanting to engage large human groups, have had little choice but to use polls, surveys, online forums, and other mechanisms that lose the interactive benefits of real-time conversational deliberation. This has been a barrier for many applications, from market research to participatory democracy. In this paper, we describe and test a novel solution that is inspired by swarm-based methods in the field of Collective Intelligence.

Collective Intelligence (CI) refers to the field of research in which the knowledge, wisdom, and insights of human groups is collected and processed to achieve more accurate insights than individuals could produce on their own [1]. Common methods for tapping the intelligence of human groups are based largely on votes, polls, and surveys of various forms. These methods have been modernized in recent years through online upvoting systems, prediction markets, and sentiment extraction via Natural Language Processing (NLP), but still rely heavily on collecting isolated responses from individual participants and aggregating those responses into statistical models that often mask important beliefs or attitudes.

A new CI method called Artificial Swarm Intelligence (ASI) has been developed in recent years and found to outperform common statistical CI methods in many scenarios, especially when populations harbor diverse and/or conflicting views [2-7]. Unlike traditional methods for groupwise intelligence that aggregate statistically, ASI is based on the decision-making dynamics of biological swarms. It works by enabling networked human groups (connected over the internet) to deliberate in real-time systems (i.e., *swarms*) that can explore a set of options in synchrony and quickly converge on solutions the group can best agree upon [2-7]. The technology of ASI was first proposed by Rosenberg in 2015 and uses intelligence algorithms modeled on the emergent behaviors of schooling fish, flocking birds, and swarming bees [2,3].

Biologists refer to the emergent decision-making abilities of natural systems as *Swarm Intelligence* because it enables many biological groups to function as "super-organisms" that can make more effective groupwise decisions than the individual members could make on their own [1, 8, 9]. ASI has been shown to enable networked human groups to achieve similar benefits, amplifying collective accuracy, insights, and cohesion through real-time dynamic interactions [10-11].

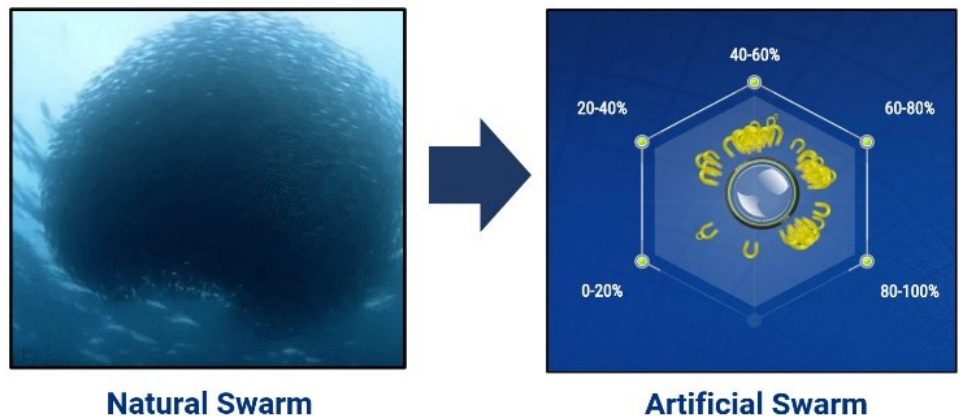

**Fig. 1.** Natural Swarm vs Artificial Swarm (Swarm® platform shown) [6]

As shown in Figure 1 above, current methods for Artificial Swarm Intelligence involve use-cases where human groups deliberate among predefined sets of options and collectively rate, rank, or select among them. For example, ASI can be used for (a) forecasting the most likely outcome from a set of possible outcomes, (b) prioritizing sets of options into ranked lists that optimizes group satisfaction, or (c) rating the relative strengths of various options against specific metrics [4-6]. While these capabilities are useful in many real-world applications, there is a need for more flexible methods of deploying ASI, especially for enabling large human groups to engage in real-time conversational deliberation on open-ended questions in which a specific set of potential solutions is not known in advance.

In the following sections, we will describe an new approach for connecting networked human groups into real-time ASI systems, leveraging the intelligence amplification benefits of Swarm Intelligence via flexible conversational interactions.

## 2    Conversational Swarm Intelligence

To address the limitations of prior CI systems modeled on the principles of Swarm Intelligence, a new architecture called Conversational Swarm Intelligence (CSI) has been developed, deployed, and tested among groups of human participants. The motivation for the CSI architecture is to enable large human groups to deliberate conversationally as real-time dynamic systems (i.e., swarms) that converge organically on solutions that maximize group satisfaction, conviction, or accuracy. By enabling the swarming process to occur conversationally, we can eliminate the need for predefined options. Instead, options emerge organically as participants suggest and debate ideas in real-time.

Current ASI systems typically engage groups ranging from 25 to 250 real-time participants. The goal of CSI is to enable conversational deliberation among similar sized groups or larger. This poses a unique challenge for an online conversational system. For example, we could bring 250 people into a single chatroom but that would not yield meaningful dialog or insight. That's because conversational quality degrades with group size [12]. Sometimes referred to as the "many minds problem," when groups grow beyond a handful of people, the conversational dynamics fall apart, providing less "airtime" per person, disrupting turn-taking dynamics, providing less feedback per comment, and reducing engagement as participants feel less social pressure to participate. In fact, putting 250 people in a chatroom would not yield a "conversation" but just a stream of singular comments with little interaction among them.

Fish schools, on the other hand, can hold "conversations" among hundreds or thousands of members with no central authority mediating the process. Each fish communicates with others using a unique organ called a "lateral line" that senses pressure changes caused by neighboring fish as they adjust speed and direction with varying levels of conviction. The number of neighbors that a given fish pays attention to varies from species to species, but it's always a small subset of the group. And because each fish reacts to an overlapping subset of other fish, information quickly propagates across the full population, enabling a single Swarm Intelligence to emerge that rapidly converges on unified decisions [13]. This is a powerful biological solution that was first emulated in networked human groups in 2021 using a technique called "hyperswarms" that was shown to enable real-time information propagation and solution convergence across a network of overlapping groups [14].

Researchers at Unanimous AI have recently applied this method to real-time human conversations via online chat. We call this a "Hyperchat" structure and it's modeled on schooling fish. In this method, a group of 300 online participants can be broken up into a large number of smaller subgroups, each ideally sized for real-time deliberations. For example, 300 participants can be divided into 60 groups of 5, with members of each subgroup automatically routed into their own unique chat room and asked to discuss an issue in parallel with the other subgroups. The subdivision alone does not foster a Swarm Intelligence because information cannot propagate across the population. We solved this through the novel use of conversational AI agents powered by Large Language Models (LLMs) to emulate the function of the lateral line organ in fish.

In particular, we insert an AI agent (referred to herein as an Observer Agent) into each of the 60 chat rooms and task that agent with monitoring the dialog in that unique room, distilling the salient content, and expressing the content in a neighboring room through natural first-person dialog. In this way, each of the 60 groups is given a sixth member that happens to be an AI observer that is tasked with conversationally expressing the real-time insights observed in one group into another group, thereby enabling information to propagate across the full "swarm."

This creates a integrated real-time system in which 300 or 3000 or even 30,000 people could hold a unified conversation on a single topic, sharing ideas, debating alternatives, and converging on global solutions that optimize groupwise support. In this way, CSI enables large populations to interact conversationally in real-time while ensuring that (a) individual members can have meaningful discourse in small deliberative groups and (b) information propagates globally leveraging the power of Swarm Intelligence. An example structure for CSI is shown in Figure 2.

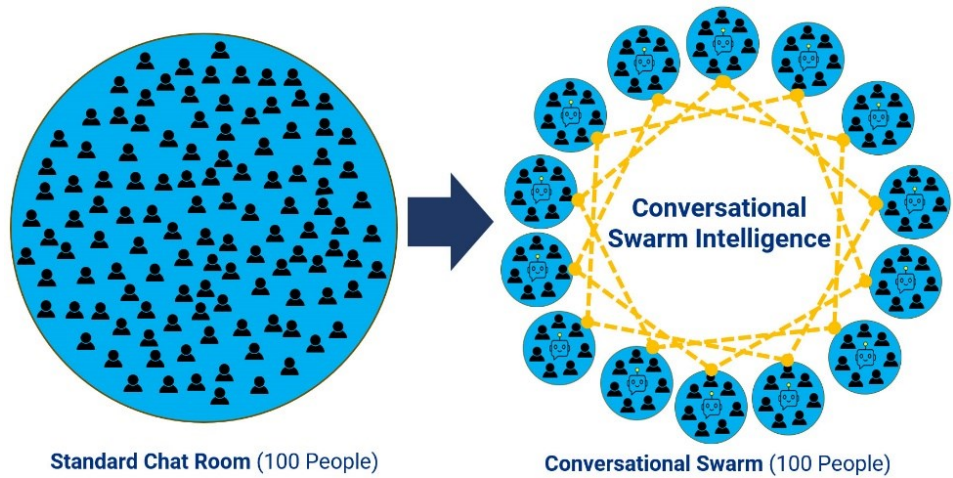

**Fig. 2.** Standard Chat versus a Conversational Swarm Intelligence structure

This method has two significant benefits over prior ASI systems. First, it enables the use of open-ended questions, empowering participants to suggest and debate options that are not pre-defined. Second, it allows users to not only indicate which options they prefer but also discuss why they prefer them. In this way, the CSI method can not only elicit solutions that maximize collective support but also capture the reasons why the group supports those solutions. In addition, the structure shown in Figure 2 above reduces the Social Influence Bias that hinders many real-time CI methods such as online discussion forums, large chat rooms, and sequential upvoting systems.

For example, when a user posts an idea, comment, or criticism into a standard forum or chat room, they can bias the full population [15]. This means that comments posted early have an overweight effect in sequential systems. It also means that a small number of strong personalities can overly influence parallel systems. We expect the CSI structure to mitigate social influence bias using the unique structure shown above because (a) each individual is only influenced by a small number of others in real-time, and (b) ideas only propagate organically after they gain momentum local.

Overall, Conversational Swarm Intelligence is designed to provide the quantitative benefits of large-scale polling, the qualitative benefits of intimate focus groups, and the intelligence amplification of swarm-based ASI systems. Considering these unique advantages, we expect CSI to be useful in a wide range of applications from market research and organizational decision-making, to collaborative forecasting, political insights, and deliberative democracy. For example, CSI could be groundbreaking for Deliberative Civic Engagement, as it can enable thoughtful conversations among representative samples of the public while facilitating convergence on solutions that maximize support across conflicting perspectives, interests, or objectives.

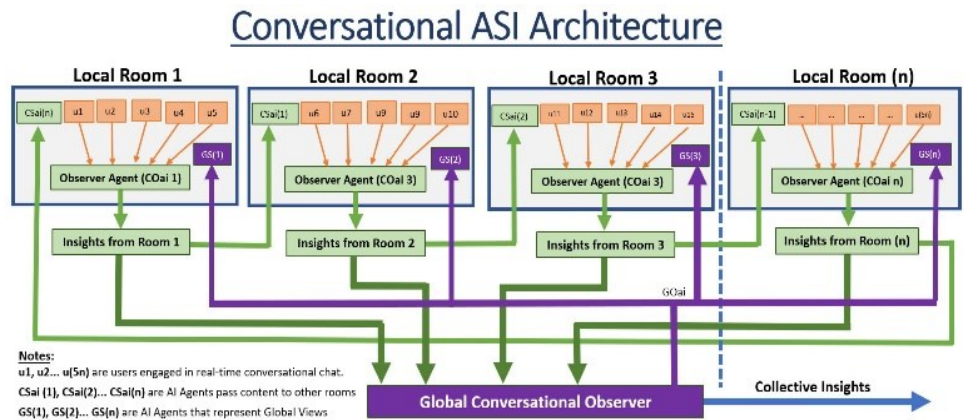

**Fig. 3.** Conversational Swarm Intelligence Architecture

As shown in figure 3 above, a novel architecture was developed to enable real-time conversational swarms among human users connected across standard computer networks. As shown, a full population (p) is broken up into (n) local rooms, each local room populated with approximately (p/n) users. Communication research suggests that ideal deliberative groups for real-time conversation range in size from 4 to 7 members, as studies show that conversations transition from authentic *dialog* to sequential *monologue* as groups approach 10 members [16]. Other research suggests maximum satisfaction for participants occurs around 5 members [17]. To optimize deliberative conversations in local groups, our current CSI heuristics aim to choose (n) such that (p/n) falls in the ideal range of 4 to 7 members.

Each local room is also provided with an Observer Agent (COai) that is enabled by making API calls to an LLM model. The AI agent was tasked with processing segments of conversational dialog within its assigned local room at regular intervals (either by elapsed time or dialog). In this study, the interval was set to 45 to 65 seconds. The processing step performs a set of functions on each block of dialog, which include – (i) identifying newly proposed suggestions related to the topic at hand, (ii) identifying comments made in support of prior suggestions, (iii) identifying comments made in opposition of prior suggestions, and

(iv) estimating the conviction level expressed when a comment either proposes, supports, or opposes a suggestion. In addition, the Observer Agent passes the summary and assessment to a databasing process that stores and aggregates suggestions and tracks groupwise conviction.

Each local room (referred to as a "ThinkTank" in the conducted studies) is provided with a Surrogate Agent (CSai) that is tasked with conversationally expressing, at intervals (measured in elapsed time and/or elapsed conversation), the local summary that was captured from a neighboring room, into an alternate room for which the agent is associated. The conversational summary is ideally expressed as conversational dialog in the first person, as if the LLM-powered Surrogate Agent is participating a member of that local group. The conversational contribution of the surrogate can be overt as to its role as an observer, for example: *"I've been observing ThinkTank 3, and they believe that job loss is a more important factor than misinformation when considering the risks of generative media."* Alternatively, the contribution of the surrogate can be configured to be more natural in its participation, expressing the dialog as if it's a first-person insight – *"From my perspective, job loss is a more important factor than misinformation when considering the risks of generative media."* In either case, the human participants are fully informed at the beginning that surrogate agents generate their conversational contributions based on the collective comments observed in alternate ThinkTanks.

Early testing of the CSI structure was conducted at Carnegie Mellon University with deliberating groups of up to 25 participants compared to standard chat rooms. Even at this relatively small population size, these tests showed significant benefit, with participants in the CSI structure producing 30% more contributions ($p<0.05$) than those using a standard chat room and 7.2% less variance, indicating that users contributed more content and participated more evenly when using CSI [24]. Based on promising results of these preliminary tests, a larger study was conducted using roughly double the number of participants (48 users in real-time deliberation) as described in Section 2 below.

## 3      Experimental Study

A cloud-based platform called Thinkscape™ was developed by researchers at Unanimous AI using the CSI architecture described above. A study was conducted to evaluate the effectiveness of conversational swarms in facilitating deliberative dialog and consensus judgements in groups of 48 users. Two separate groups of 48 users were convened in the Thinkscape platform at different times, and each of the two sessions responded to the same two questions in the same order: *"Which jobs are most at risk from AI automation in the next ten years?"* and then *"What's the biggest risk we face from AI in the next ten years?"* All users were paid participants recruited from a professional sample provider. No user participated in both sessions. In the first session, a standard chat room was used to answer the first question, and a conversational swarm was used to answer the second question. In the second session, this assignment was reversed: the group first used the conversational swarm to answer the jobs question, and then used standard chat. In this way, we were able to measure the effect of the chat structure on deliberative outcomes while controlling for the question being asked.

For each question, the group of 48 participants was given six minutes to deliberate via chat and reach a collective consensus for a single preferred answer. The consensus was determined in all cases using a preference-labeling algorithm. This preference-labeling algorithm worked as follows: after every 5 messages in each room, or at most 15 seconds after a new message appeared in that room, API calls to a Large Language Model (in this case GPT 3.5) were automatically generated, causing the LLM to identify the answers that each person in that room preferred in response to the question being discussed. The strength of the individual preferences assessed were normalized on a scale of -3 (extreme negative,) to +3 (extreme positive). This preference-labeling architecture was manually validated for accuracy of the preference labels and was deemed highly accurate at labeling user preference during real-time conversations in the Thinkscape platform. If a user did not mention an item or expressed no preference, their preference for that item was set to 0. The net preference for each item was calculated as the average preference for that item over all individuals. The item that showed the highest net preference at the end of the question was selected as the final answer.

## 4 Results

To quantify the change in behavior of groups when working together in conversational swarms as compared to standard chat rooms, the number of messages sent by each participant during each deliberative session and the total informational content sent by each participant (measured in number of characters) was calculated. A paired t-test was used to measure the significance of differences in conversational contributions between the chat room and the conversational swarms. As shown in Table 1 below, participants in conversational swarms produced 46% more messages (0.699 vs 1.021 messages per minute, $p<0.001$) compared to the standard chat room, and 51% more characters in their chat messages (32 vs 49 characters per minute, $p<0.001$).

**Table 1.** User Contribution in Conversational Swarms and Standard Chat

| Measurement | Metric | Collaboration Structure | |
| --- | --- | --- | --- |
| | | Single Chat Room | Conversational Swarm |
| Messages per Minute | Mean | 0.699 | 1.021 |
| | Variance | 0.162 | 0.226 |
| Characters Per Minute | Mean | 32.4 | 49.0 |
| | Variance | 868.7 | 1203.4 |

The number of messages and number of characters each user contributed per minute in the single chat room and conversational swarm are compared in Figures 4 and 5 below. The overwhelming majority of users contributed more using CSI as compared to the standard chat room, as measured in both messages per minute and characters per minute. Interestingly, this effect applies to all users, but more so for users who participated the least: these users increased their message frequency and character rate more than users who were already vocal participants in the single chat room.

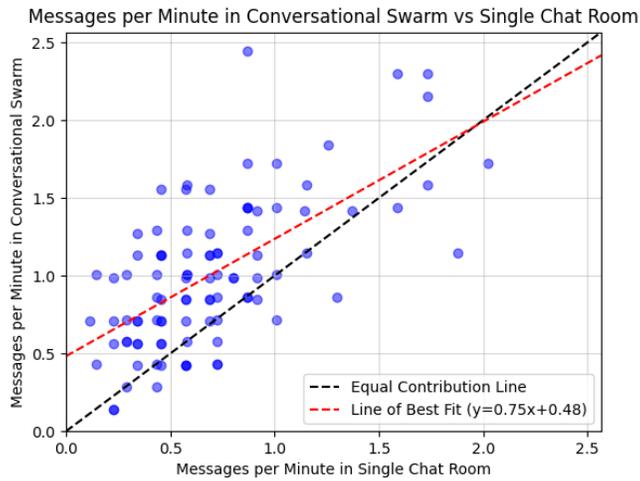

**Fig. 4.** Messages per minute by user in the Conversational Swarm and Chat Room

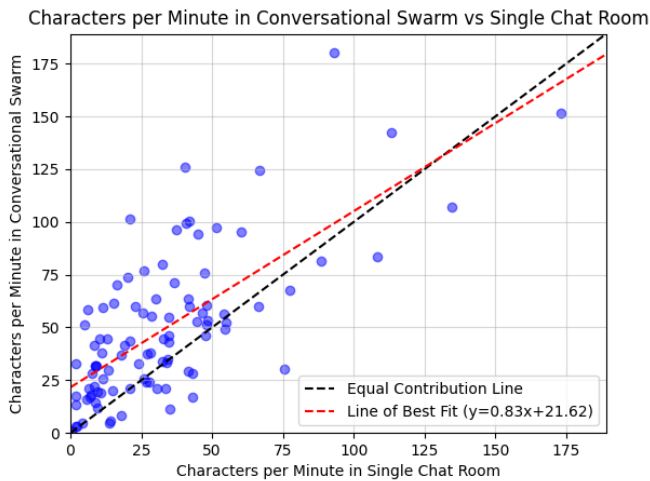

**Fig. 5.** Characters per minute by user in the Conversational Swarm and Chat Room

In addition to the average contribution per user increasing when using CSI, the relative difference between the heaviest and lightest contributors decreases as well. The boxplots in Figures 6 and 7 show the distribution of contribution rates for each of the two structures. Here, the quartiles of contribution rates are shown as large boxes and the 10$^{th}$ and 90$^{th}$ percentiles are shown as thinner boxes. These charts reveal that the overall contribution increases from Standard Chat to Conversational Swarm, with the median contributions in the Conversational Swarm exceeding the 75$^{th}$ percentile in the Single Chat Room for both messages per minute and characters per minute.

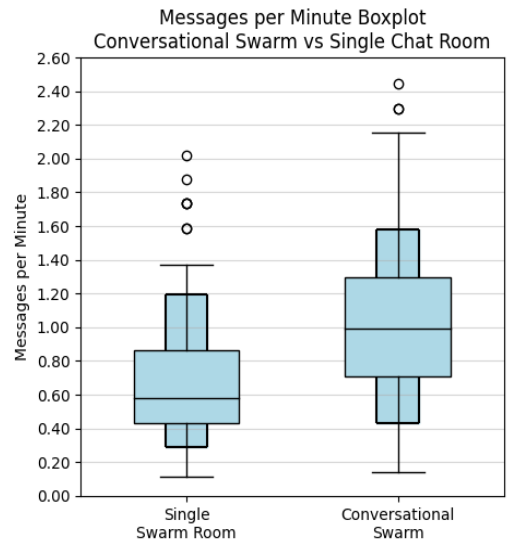

**Fig. 6.** Boxplot of user messages per minute in Standard Chat vs CSI.

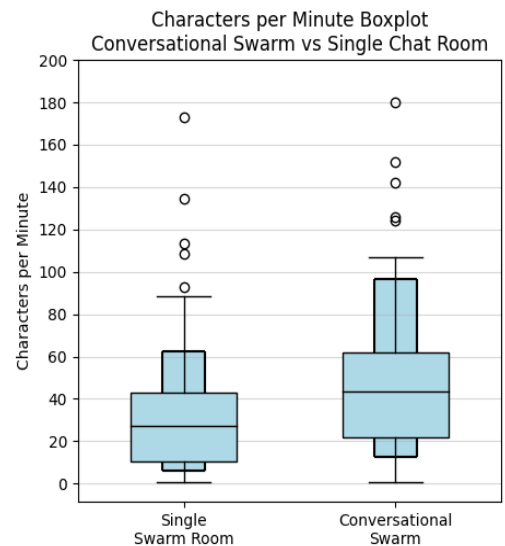

**Fig. 7.** Boxplot of user characters per minute in Standard Chat vs CSI

To quantify the degree to which the discussion in each room was dominated by a small minority of vocal participants, we calculated the ratio of contribution for the 90[th] vs 10[th] percentiles of participants. We call this the *contribution ratio*. As shown in Table 2 below, the contribution ratio is 13% higher in the standard chat room as compared to the CSI system when measuring contribution in messages-per-minute. The contribution ratio was 37% higher when measuring the contribution in characters-per-minute. This suggests that the CSI structure allowed deliberations to be dominated less by a small number of vocal individuals, thereby fostering more equal contributions from participants compared to a standard chat environment.

**Table 2.** Comparison between 10$^{th}$ and 90$^{th}$ Percentile Messages per Minute Contributors in Conversational Swarms and Standard Chat

| Measurement | Metric | Collaboration Structure | |
| --- | --- | --- | --- |
| | | Chat Room | Conversational Swarm |
| Messages Per Minute | 10$^{th}$ Percentile | 0.289 | 0.431 |
| | 90$^{th}$ Percentile | 1.196 | 1.581 |
| | Contribution Ratio | 4.14 | **3.67** |
| Characters Per Minute | 10$^{th}$ Percentile | 6.0 | 12.6 |
| | 90$^{th}$ Percentile | 62.6 | 96.6 |
| | Contribution Ratio | 10.4 | **7.6** |

After Session 2 completed, participants in this session were asked to answer a survey that investigated three questions: (i) In which structure did they prefer working, (ii) In which structure did they feel more heard, and (iii) In which structure did they feel the group generated better justifications for their answers? All of these questions were asked in a 5-point scale, including the five following answers: Single chat room by a lot, Single chat room by a little, No preference, Conversational swarm by a little, and Conversational swarm by a lot. All users who participated in the session completed the survey. Binomial statistical tests were performed on this survey data that compared the fraction of users who responded in favor of conversational swarms as compared to all other answers (including both those that favored the single chat room and those that expressed no preference).

The results of these three survey questions are shown in figures 8-10. The group significantly preferred the conversational swarm structure (65% in favor, 18% no preference, 16% opposed, $p<0.02$), felt significantly more heard in the conversational swarm (67% in favor, 24% no preference, 8% opposed, $p<0.005$), and felt the group generated better justifications for their answers (51% in favor, 29% no preference, 20% opposed, $p<0.5$).

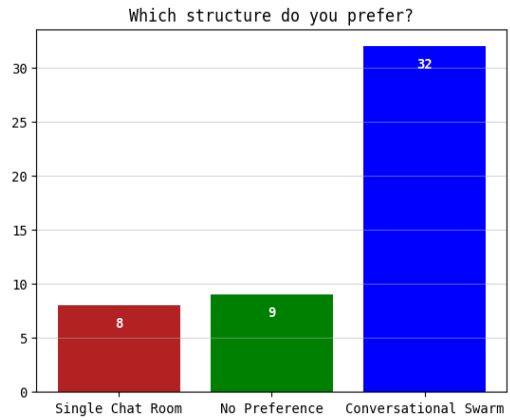

**Fig. 8.** Distribution of responses for: "Which structure did you prefer?"

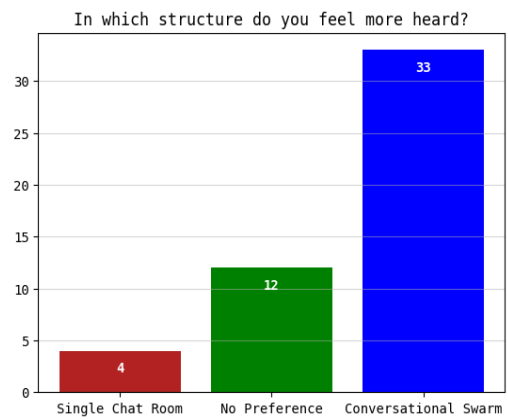

**Fig. 9.** Distribution of responses for: "In which structure did you feel more heard?"

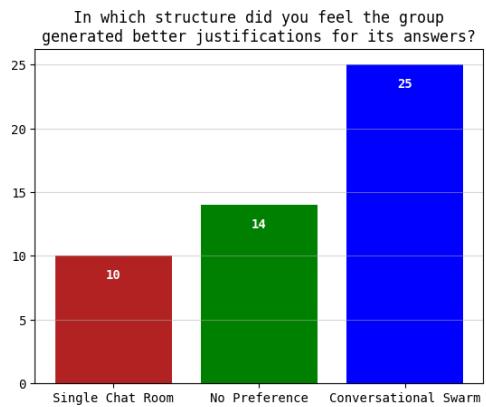

**Fig. 10.** Distribution of responses for the question: "In which structure did you feel the group generated better justifications for its answers?"

## 5      Conclusions

We introduce Conversational Swarm Intelligence, a novel technology that allows large, networked groups to hold coherent real-time deliberations via online chat and reach consensus. We performed a human study with 48-member groups, comparing their ability to deliberate in a CSI structure vs a standard chat room.  The results showed that deliberating participants generated 46% more messages ($p<0.001$) and 51% more content ($p<0.001$) when using CSI as compared to a standard chat room.

In addition, we observed that conversational contributions in CSI were more evenly balanced, while conversations in standard chat rooms were dominated by a small number of users. This was quantified, showing a 37% decrease in the contribution ratio between the 90th percentile contributors and the 10th percentile contributors (by characters per minute). We attribute these two benefits (more participation overall and more balanced contributions) to the effect of individuals interacting in smaller rooms while still connected to the larger population through conversational AI agents. We expect these benefits to increase with larger populations of participants, which future work will validate.

We also observed that the group significantly preferred working in Conversational Swarms as compared to large single groups ($p<0.05$), and that individuals felt their contributions had more impact in conversational swarms ($p<0.01$). These results suggest that CSI is a promising technology for large-scale conversational deliberations, increasing participant engagement, contribution, and satisfaction.

Future work will explore the ability of Conversational Swarm Intelligence to amplify the groupwise accuracy of collective forecasting and decision-making, the effectiveness of groupwise brainstorming and ideation, and the representativeness of groupwise assessments and evaluations. Future work will also strive to test CSI in real-time deliberative democracy and civic engagement contexts and will evaluate whether the higher engagement seen in CSI can translate into more accurate and actionable outcomes. And finally, future work will explore significantly larger groups of participants, extending to real-time deliberations among hundreds or thousands of simultaneous participants, and will deploy the CSI architecture for use in real-time voice chat, video chat, and VR chat environments.